\documentclass{emulateapj}
  \usepackage{natbib}
\usepackage{graphics}
  \newcommand{\bq}{\begin{equation}}
  \newcommand{\eq}{\end{equation}}
  \newcommand{\ba}{\begin{eqnarray}}
  \newcommand{\ea}{\end{eqnarray}}

  \def\mag{{\rm mag}}

\begin{document}

  \title{Cosmology with Photometric Surveys of Type Ia Supernovae}

\author{Yan Gong$^{1,2,3}$}
\author{Asantha Cooray$^{1}$}
\author{Xuelei Chen$^{2,4}$}
\affil{$^1$Department of Physics \& Astronomy, University of California, Irvine, CA 92697}
\affil{$^2$National Astronomical Observatories, Chinese Academy of Sciences, Beijing, 100012, China}
\affil{$^3$Graduate School of Chinese Academy of Sciences, Beijing 100049, China}
\affil{$^4$Center of High Energy Physics, Peking University, Beijing 100871, China}

  \begin{abstract}
We discuss the extent to which photometric measurements alone can be
used to identify Type Ia supernovae (SNIa) and to determine the
redshift and other parameters of interest for cosmological studies.
We fit the light curve data of the type expected from a survey such
as the one planned with the Large Synoptic Survey Telescope (LSST)
and also to remove the contamination from the  core-collapse
supernovae to SNIa samples. We generate 1000 SNIa mock flux data for
each of the LSST filters based on existing design parameters, then
use a Markov Chain Monte-Carlo (MCMC) analysis to fit for the
redshift, apparent magnitude, stretch factor and the phase of the
SNIa. We find that the model fitting  works adequately well when the
true SNe redshift is below 0.5, while at $z < 0.2$ the accuracy of
the photometric data is almost comparable with spectroscopic
measurements of the same sample. We discuss the contamination of
Type Ib/c (SNIb/c)  and Type II supernova (SNII) on the SNIa data
set. We find it is easy to distinguish the SNII through the large
$\chi^2$ mismatch when fitting to photometric data with Ia light
curves. This is not the case for SNIb/c. We implement a statistical
method based on the Bayesian estimation  in order to statistically
reduce the contamination from SNIb/c for cosmological parameter
measurements from the whole SNe sample. The proposed statistical
method also evaluate the fraction of the SNIa in the total SNe data
set, which provides a valuable guide  to establish the degree of
contamination.

  \end{abstract}


  \maketitle

  \section{Introduction}

The cosmological applications of luminosity-distance measurements to
Type Ia supernovae (SNeIa) are now well known
\citep{riess98,perlmutter99,leibundgut01}. While the current sample
of SNeIa-based distances are limited to a few hundred SNe
\citep{astier06,wood-vasey07,kowalski08,hicken09,kessler09}, future
surveys are now planned to increase the sample size to a few
thousand or more that could potentially allow a few percent accurate
dark energy equation of state measurements in several redshift bins
between $0 < z < 1$ (see e.g. \citealt{howell09,sarkar08}). The main
challenge for constructing large samples are likely to be
spectroscopic follow-up measurements to identify if each supernova
detected in a photometric monitoring campaign is Type Ia and to
establish the redshift of that supernova.

In addition to the planned space-based programs such as the Joint Dark
Energy Mission (JDEM) \footnote{http://jdem.gsfc.nasa.gov/}, in the
near future, there will also be several ground based photometric surveys
for cosmological measurements and other astronomical studies. These
include the Dark Energy Survey
(DES)\footnote{http://www.darkenergysurvey.org/}, the Pan-Starrs
survey \footnote{http://pan-starrs.ifa.hawaii.edu}, and ultimately
the Large Synoptic Survey Telescope (LSST)
\footnote{http://www.lsst.org/}, which plans to monitor a large area
of the sky every few days leading to a large sample of transient
sources including supernovae. Given the large size of the samples of
SNe expected, it is highly unlikely to have spectroscopic follow-ups
for all or even a large fraction of them. Due to this limitation it
appears challenging to obtain cosmological measurements with the
SNIa seen by LSST. Since LSST is likely to detect a few hundred
thousand or more SNe per year, it would be highly desirable to
identify whether a given SN as Type Ia or not, and to extract useful
parameters such as redshift and luminosity with photometric data
alone. If reliable techniques could be established, then even with a
large degradation in accuracy for individual data compared with the
case where spectroscopic data are also available, given the large
number statistics expected, one could still aim to achieve a good
measurement of cosmological parameters.

In this spirit we pursue a study to establish the extent to which
photometric data from a survey like LSST can be used to identify
SNeIa and to measure the cosmological parameters. We do this by
fitting the photometric light curve data with sampling and errors
consistent with LSST. Our mock SNe samples also include
core-collapse supernovae and we vary the fractions expected based on
the current rate estimates of various types of SNe. Our MCMC
analysis are focused on a joint parameter estimation including the
redshift, apparent magnitude, stretch factor and the phase of the
SNIa. We find that the model fitting  works adequately well when the
true SNe redshift is below 0.5. At $z < 0.2$, photometric data of
the type expected with LSST provide an accuracy that is close to the
case when spectroscopic measurements are also available, with the
redshift determined separately from spectroscopic data leading to
one less parameter in MCMC fits than photometric light curves.

We also focus on the contamination of  Type Ib/c (SNIb/c)  and Type
II supernova (SNII) on the SNIa data set. We find it is easy to
distinguish the SNII from SNIa's through the large $\chi^2$ mismatch
in the fitting to Ia light curves. This is not the case for SNIb/c
and they provide the main contamination to Ia measurements. In
addition to a cut in $\chi^2$ values, we implement a statistical
Bayesian estimation method to  reduce the contamination from SNIb/c
in the subset of SNe sample selected for cosmological measurements.
This technique also establishes statistically  the fraction of the
SNIa in the total SNe data set, which provides a valuable guide  to
the degree of contamination from Ib/c's.

We employ the filter functions as currently publicized by the LSST
team in addition to survey parameters outlined in Ivezi\'c et al.
(2008). We note that while our work is focused towards a survey like
LSST, others have also consider the use of photometric data alone
for SNe distance measurements
\citep{johnson06,sullivan06,kuznetsova07,wang07,kim07,zentner09}.

The discussion is organized as follows. In the next Section we
describe our procedure to simulate SNeIa data in a survey like LSST
and move on to discuss our six parameter model fits to the
multi-wavelength light curves from a large mock sample using a MCMC
analysis. In Section~4, we discuss the contamination from Type II
and Ib/c SNe to Ia photometric samples and a way to statistically
reduce the contamination from Ib/c's using a technique that
implements the Bayes theorem.

  \section{Simulating SNIa Observed Flux data}

  In this Section, we describe the process to generate the various
SNe data. We first discuss the observed SNIa mock flux data.

  \subsection{The Mock Light Curve}

  The apparent observed flux from a supernova at $z$ can be written as
the convolution of the spectral energy distribution (SED) and the
transmission function of the telescope,
  \bq
  f_{obs} = \int T_X(\lambda_{obs}){\rm SED}(\lambda_{obs},
  t_{obs}, s, E_{B-V}^{host}, E_{B-V}^{MW})d\lambda_{obs},
  \eq
where $T_X(\lambda_{obs})$ is the filter response for band $X$,
$\lambda_{obs}$ is the observed wavelength, $t_{obs}$ is the
observation date, $s$ is the stretch factor, and $E_{B-V}^{host}$
and $E_{B-V}^{MW}$ are the color excess for the host galaxy and the
Milky Way respectively.

  \begin{figure}[htbp]
\begin{center}
  \includegraphics[scale = 0.34]{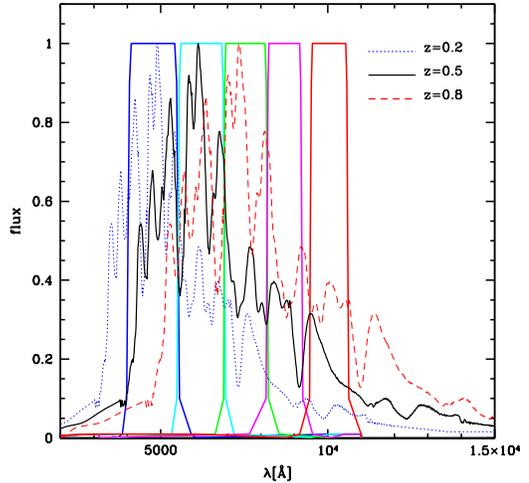}
\end{center}
  \caption{\label{fig:filter} The LSST filters used in our analysis.
  From left to right are g, r, i, z and y-band functions. The Type Ia SEDs
  of SNe for $z=0.2$, $z=0.5$ and $z=0.8$ at the day of rest-frame B
  band peak magnitude are also shown. The flux
  is in an arbitrary unit.}
  \end{figure}

  The transmission functions used in our analysis for 5-bands of LSST
are shown in Fig.\ref{fig:filter} \citep{lsst_filter}.\footnote{We
note that there are several filter designs for LSST including a
scenario involving 6 filters. Here, we focus on the 5-band case with
simple filters.} We also plot rest-frame SEDs for Type Ia SNe at
$z=0.2$, 0.5, and 0.8. These SED templates are from {Nugent et al.
(2002)} and they cover the spectral wavelength from $1000$ to
$25000{\rm \AA}$ in rest-frame days from $-20$ to $70$ with respect
to the B-band maximum light day. The SNe flux for the epochs before
-20 are set to be zero.

There are now several techniques to parameterize the SNIa light
curves, such as the 15-day decline after the B-band maximum light
$\Delta m_{15}$ \citep{phillips93} and the multicolor light curve
shape method (MLCS and the update version MLCS2k2)
\citep{riess96,jha07}. In this paper, we calibrate the SNIa light
curve with the time-scale and stretch factor relation following the
works of \citet{perlmutter97,perlmutter99}. By stretching and
compressing the time axis around the rest frame B-band maximum light
day, this method can fit the observed light curve very well using
the light curve template \citep{goldhaber01}.  Then this SED can be
re-scaled by the apparent, unextincted B-band peak magnitude
  \bq
  m_B = M_B+5{\rm log_{10}}d_L(z,{\bf \theta})+25-\alpha(s-1)+\Delta m,
  \eq
where $M_B$ is the B-band absolute peak magnitude, $d_L$ is the
luminosity distance which is a function of the redshift $z$ and a
broad set of cosmological parameters denoted by ${\bf \theta}$ and
$\alpha$ is the coefficient of the relation between $s$ and $m_B$.
Here we take $M_B=-19.3$, $\alpha=1.5$ \citep{knop03,astier06}, and
the set of cosmological parameters ${\bf \theta}$ with
$\Omega_{m0}=0.27$, $\Omega_{\Lambda 0}=0.73$ and $h_0=0.71$
\citep{komatsu09} where $\Omega_{m0}$ and $\Omega_{\Lambda 0}$ have
the usual meaning with the present-day matter and dark energy
density parameters and $h_0$ is the dimensionless Hubble constant.
Besides, we also consider the dispersion $\Delta m$ of the
rest-frame B-band peak magnitude after the calibration of the
stretch factor. The B-band filter we use is from the Johnson-Morgan
system \citep{bessell90,bessell05}. Also, the time scale of the SED
is calibrated by the stretch factor $s$, which is assumed to be
available from $-15$ to $35$ around the B-band maximum luminosity
day \citep{astier06}. We note that there is also an intrinsic color
scatter, $\sigma_{int}^{B-V}$ (standard deviation), that should be
taken into account when producing mock light curves. Based on prior
work, we find this uncertainty to be small with a value of
$\sim0.05$ mag \citep{phillips99,jha07}. Hence it would not affect
our results much, so that for simplicity we don't consider it here.

During the transit, the supernovae light will be partly absorbed by
the  dust of the host galaxy. We employ the reddening law of
{Cardelli et al. (1989)} with $R_V=3.1$, from infrared to
far-ultraviolet ($0.3\mu m^{-1}\le x \le 10\mu m^{-1}$, where
$x=1/\lambda$). For the optical to near ultraviolet wavelength range
($1.1\mu m^{-1}\le x \le 3.3\mu m^{-1}$), we use an updated version
for extinction given by {O'Donnell (1994)}. The latter uses the same
analytical form for extinction as {Cardelli et al. (1989)}  but with
values of the fitting parameters revised slightly from the previous
version.  The level of extinction we assume here is consistent with
the one measured recently by {Menard et al. (2009)} corresponding to
large angular scales based on galaxy-QSO cross-correlation in SDSS.

  Since the SNe are at a different redshift than the observer, the
spectrum is redshifted for both wavelength and the phase, i.e.
$\lambda'=\lambda(1+z)$ and $t'=t(1+z)$. We also apply an
extinction associated with dust in the Milky Way
\citep{burstein82,schlegel98}, and assume that we have a perfect
measurement of $E_{B-V}^{MW}$. This assumption has no effect
on our final conclusions.  While the extinction of the Milky
Way have different values for different sky regions, we do not have
any information on the exact field selection of future SNe surveys
from ground. Thus, we do not account for sky variation of extinction
and simplify by just taking an average value with $E_{B-V}^{MW}
\approx 0.03$ and $R_V=3.1$ for the extinction law. Finally, the
spectrum is integrated with the LSST filters to get the mock light
curve sampling in each of the 5 LSST filters.

  Since the mean redshift of the SNe detections with LSST main survey
is expected to be about $0.5$, and the deeper, but smaller, survey
can potentially detect SNe out to $\sim 1$, we choose the redshift
range from 0.01 to 1.1 when making mock SNe samples.

When creating large samples, we assume the flat $\rm \Lambda CDM$
model with $\Omega_{m0}=0.27$ and $h_0=0.71$. Then, the redshift
$z$, stretch factor $s$, the extinction of the host galaxy
$E_{B-V}^{host}$ and the magnitude dispersion $\Delta m$ are
generated from the Gaussian distribution with truncated tails as
follow: $0.01\le z\le 1.1$ with $\bar{z}=0.5$ and $\sigma_z=0.4$,
$0.6\le s\le 1.4$ with $\bar{s}=1$ and $\sigma_s=0.3$, $-0.1\le
E_{B-V}^{host}\le 0.3$ with $\overline{E_{B-V}^{host}}=0.0$ and
$\sigma_{E}=0.2$ and $-0.3\le\Delta m\le0.3$ with $\overline{\Delta
m}=0.0$ and $\sigma_{\Delta m}=0.17$. This extra dispersion acts as
an extra source of noise in our mock data
\citep{sullivan06,hamuy95,hamuy96,phillips99,guy05}.

  \subsection{The Photometric Error and The Cadence}

  The photometric error we use for LSST comes from {Ivezic et al.
(2008)} and takes the form of
  \bq
  \sigma_{phot}^2 = \sigma_{sys}^2 + \sigma_{zero}^2 + \sigma_{rand}^2,
  \eq
where $\sigma_{sys}$ is the systematic photometric error which is
designed to be very small ($<0.005 \mag$). $\sigma_{zero}$ is the
absolute photometric error that we set to be $\sigma_{zero}=0.02
\mag$ \citep{astier06,sullivan06}. We note that, in practice,
there is only one zero-point realization in any given experiment that is
applied to all supernovae. This would result in a non-diagonal
covariance matrix for the distance modulus \citep{kim06}. However,
since the inclusion or non-inclusion of this covariance
does not change the principles of our methodology, for simplicity,
we ignore this correlation here. We do suggest that it must be considered in
an analysis of real data.

In equation~(3),  $\sigma_{rand}^2$ is the random photometric error for
point sources given by
  \bq
  \sigma_{rand}^2=(0.04-\gamma)x+\gamma x^2.
  \eq
Here $\gamma$ is a parameter related to the sky brightness and
readout noise, among others. and $x=10^{0.4(m-m_5)}$, where $m$ is
the magnitude and $m_5$ is $5\sigma$ depth for a detection of a
point source in each of LSST bands. The $m_5$ is a function of the
sky brightness, the seeing, the exposure time, atmospheric
extinction, the airmass and the overall throughput of the
instrument. All of the value of these parameter can be found in
Table~2 of {Ivezic et al. (2008)}.

  We randomly generate the first observational day from -20 to 35
rest-frame days to ensure that we always have enough data to
establish the stretch factor. We next randomly select the data point
to occur every 3 or 4 days based on the cadence of the LSST
\citep{ivezic08}. Finally, about 1000 mock SNIa flux data are
generated for each of the five filters.

\begin{figure}[htbp]
  \includegraphics[scale = 0.36]{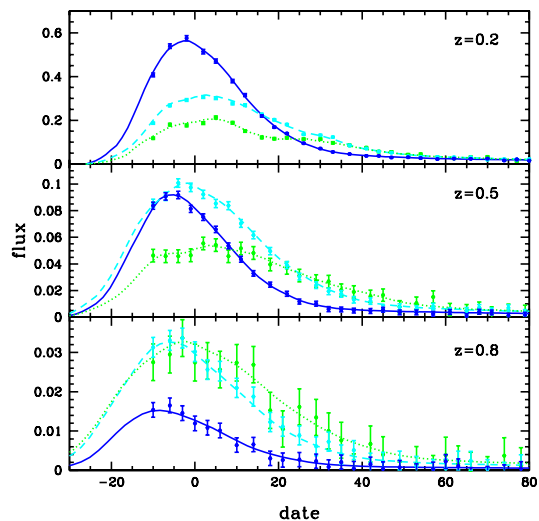}
  \caption{\label{fig:LC} The examples of LSST mock SNIa light
  curves and observational data. The solid blue, dashed cyan
  and dotted green lines are the g, r and i band light curves
  respectively, and the first observe-day is set at -10 day
  in the observer-frame. The flux is in an arbitrary unit.}
  \end{figure}

  In Fig.\ref{fig:LC}, we show the examples of the mock light curves
in g, r and i bands at different redshifts. The mock flux is created
from the Gaussian distribution with the mean on the light curve.
Here we set the first observe-day $t_0^{obs}=-10$ in the
observer-frame.

  \section{Fitting The Light Curve}

  There are six light curve parameters that we hope to extract from
multi-wavelength light curve fitting, These parameters are the $z$,
$m_B$,   $s$, $E_{B-V}^{host}$, $\Delta m$ and $t_0^{rest}$ (i.e.
the rest-frame date for the first observe-day). The $\chi^2$
statistical method is employed here with
  \bq
  \chi^2 = \sum_{i}^{t}\sum_{j}^{bands}\bigg\{ \frac{f^{obs}_{ij}-
           f^{th}_{ij}(T_j;z,m_B,s,E_{B-V}^{host},\Delta m,t_0^{rest})}
           {\sigma^{obs}_{ij}} \bigg\}^2,
  \eq
where $f_{ij}^{obs}$, $f_{ij}^{th}$, and
$\sigma_{ij}^{obs}=\sigma_{phot}$ are the observed, theoretical flux
and observed error for the observe-day $t_i$ and band $j$, and $T_j$
is the transmission of band $j$. The summation goes through all
bands and days with observed samplings of the light curves.

\subsection{The Markov Chain Monte Carlo Technique }

The best-fit value for each light curve parameter usually can
be found using the nonlinear least-squares fitting technique (e.g.
sullivan et al. 2006). Here considering the number of the
parameters, the efficiency and the accuracy, we would like to employ
the MCMC technique to perform the fitting process. This method does
not require to assume a Gaussian distribution for the likelihood,
and it is easy to perform the marginalization over other parameters
when quoting error for one parameter. Most importantly, it is very
efficient for the multi-parameter fitting
\citep{neil93,lewis02,macKay03,doran04,gong07,trotta08}.

  Our purpose is to estimate the posterior probability $P(\theta|{\bf
D})$ for the parameter set ${\bf \theta}$ given the observational
data set ${\bf D}$. Based on the Bayes theorem
  \bq
  P(\theta|{\bf D}) = \frac{\mathcal{L}({\bf D}|\theta)P(\theta)}{P({\bf D})},\label{eq:bys_th}
  \eq
where $\mathcal{L}({\bf D}|\theta)\sim e^{-\chi^2/2}$ is the
likelihood which denotes the probability to get ${\bf D}$ given the
parameters $\theta$, $P(\theta)$ is the prior probability for
$\theta$ and $P({\bf D})$ is the normalization factor which would
not affect our analysis  here.

  The Metropolis-Hastings algorithm is applied in our MCMC technique
to decide if a new point should be accepted by an acceptance
probability:
  \ba
  {\bf a}({\theta_{n+1} | \theta_n}) & = & \min
  \Bigg\{\frac{P(\theta_{n+1} | {\bf D})\; {\bf q}(\theta_n |
  \theta_{n+1})}{P(\theta_n|{\bf D})\;
  {\bf q}(\theta_{n+1}|\theta_n)}\ , 1\Bigg\} \\
  \nonumber\\& = & \min
  \Bigg\{\frac{{\mathcal{L}}({\bf D}|\theta_{n+1})\;
  {\bf q}(\theta_n|\theta_{n+1})}{{\mathcal{L}}({\bf D}|\theta_n)\;
  {\bf q}(\theta_{n+1}|\theta_n)}\ , 1\Bigg\}, \label{eq:accept}
  \ea
where ${\bf q}(\theta_{n+1}|\theta_n)$ is the proposal density to
propose a new point $\theta_{new}$ given a current point $\theta_n$
in the chain. Here we assume uniform prior probabilities for the
parameters which is canceled in Eq.(\ref{eq:accept}). If ${\bf a}=
1$, the new point $\theta_{new}$ is accepted; otherwise, the new
point is accepted with probability ${\bf a}$. This process are
repeated until a new point is accepted, and then we set
$\theta_{n+1} = \theta_{new}$. Also, we set a uniform
Gaussian-distributed proposal density for every point, so that it is
independent of the position on the chain, i.e. ${\bf
q}(\theta_{n+1}|\theta_n)={\bf q}(\theta_n|\theta_{n+1})$, we then
have
  \bq
  {\bf a}({\theta_{n+1}|\theta_n}) =
  \min\Bigg\{\frac{{\mathcal{L}}({\bf D}|\theta_{n+1})}{{\mathcal{L}}({\bf D}|\theta_n)}\
  , 1\Bigg\}.
  \eq
Since the proposal density determines the step size of the MCMC
process, it is closely related to the convergence and mixing of the
chain. Here we adopt the adaptive step size Gaussian sampler given
by \citep{doran04}. The criterion of the convergence we use was
described in \citet{gelman92}, and after convergence we freeze the
step size \citep{doran04}.

  The ranges of the parameters when we run the MCMC are set as follow:
$z\in(0,2)$, $m_B\in(10,30)$, $s\in(0.5,1.5)$,
$E_{B-V}^{host}\in(-0.5, 0.5)$, $\Delta m\in(-0.5,0.5)$ and
$t_0^{rest}\in(-20,40)$. For each mock SNIa, we take about $10000$
chain points to illustrate the probability distribution of the
parameters after the burn-in and thinning process.

\subsection{The Light Curve Fitting Results}

\begin{figure}[htbp]
  \includegraphics[scale = 0.36]{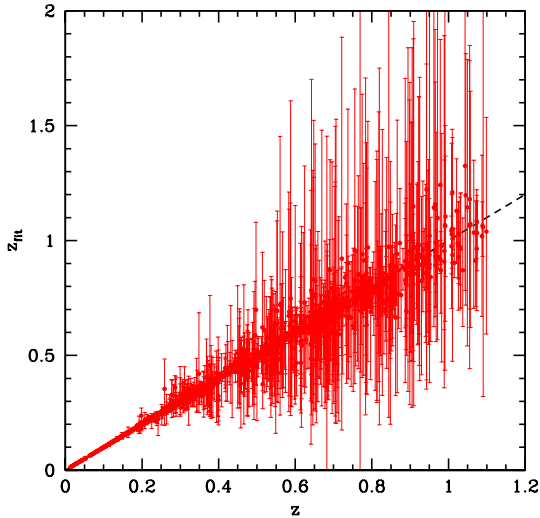}
  \caption{\label{fig:z_fit} The intrinsic redshift of each of the 1000 mock SNeIa light curves
  compared to the photometric redshift estimated with multi-parameter MCMC fits to multi-wavelength light curves.
  The best fit value and $1\sigma$ errors are  shown. The redshift estimation is remarkably
  accurate at low redshifts when $z < 0.2$
  with errors comparable to either the spectroscopic
  measurements of redshift or theoretical uncertainty in the redshift coming
  from peculiar velocities and bulk flows, among others. At $z > 0.5$, the fitted redshift errors
  are significantly larger
  because of the large photometric errors $\sigma_{phot}$.}
  \end{figure}

  In Fig.\ref{fig:z_fit}, we compare the input redshift of each of our
1000 SNeIa in the simulation with the redshift obtained from MCMC
fitting of SEDs to the multi-wavelength light curves. We find that
when $z<0.2$ the SNeIa light curves are adequately sampled with
enough accuracy to allow good redshift estimates along with other
parameters, with uncertainties as small as $0.001$. Such an error is
comparable with the spectroscopic measurements, and even if the
spectroscopic measurements could provide a higher precision on the
measurement of the redshift, in any case the unknown bulk flows
\citep{cooray06,zhang-chen08} would produce an error on the redshift
at this level. For the medium redshift $0.2<z<0.5$, the estimated
redshift is still useful but the the uncertainty is about 0.1. For
$z>0.5$, the apparent magnitude becomes large, and since
$\sigma_{phot}\sim10^{0.4m}$, the redshift errors increase quickly
with increasing redshift and can reach $\sim 1$. The limitation at
high redshift is also due to lack of near-IR photometric coverage
and addition of IR bands beyond the $z$-band will improve
photometric determinations when $z > 0.5$.

  \begin{figure}[htbp]
  \includegraphics[scale = 0.36]{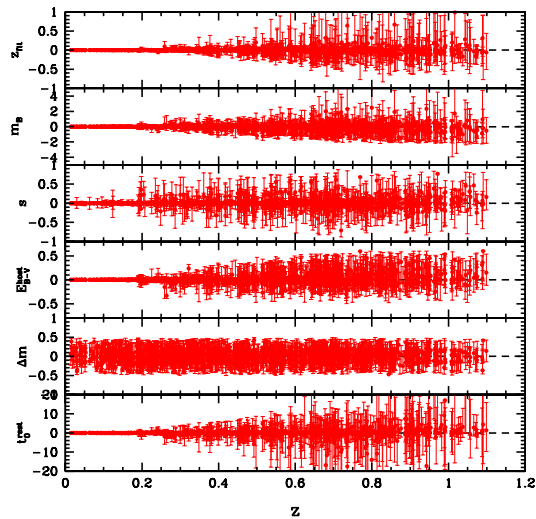}
  \caption{\label{fig:para_contr} The residuals for the
  six parameters in the MCMC analysis with $1\sigma$ errors
  for the 1000 SNeIa in the mock sample.
  The results are pretty good for $z<0.2$, except for
  $\Delta m$ since it can be seen as the noise and is
  independent on the redshift.}
  \end{figure}

\begin{figure}[htbp]
  \includegraphics[scale = 0.36]{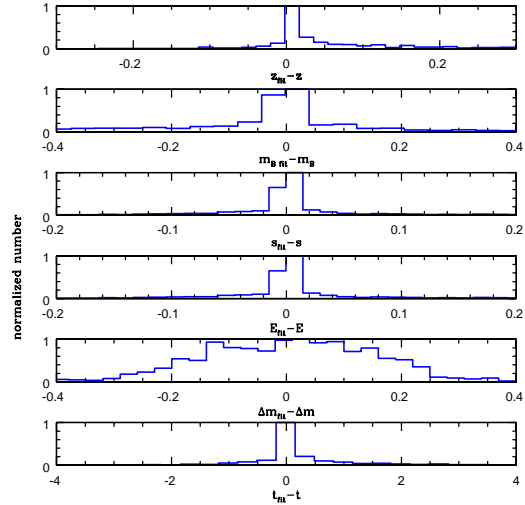}
  \caption{\label{fig:para_dis} The distribution of the fitting
  value minus the actual value for each light curve parameter.
  The number has been normalized.
  }
  \end{figure}

  The residuals and $1\sigma$ errors for the total six fitting
parameters in the MCMC analysis are shown in
Fig.\ref{fig:para_contr}. Similar to the redshift, the
multi-wavelength light curve model fitting leads to parameter
accuracies that are remarkably accurate when $z<0.2$, except for
$\Delta m$ as it acts as an extra source of noise independent of the
redshift. As shown in Fig.\ref{fig:para_dis}, over the whole
redshift range studied out to z of 1.1, the dispersion of the
fitting $z$, $m_B$, $s$, $E_{B-V}^{host}$, $\Delta m$ and
$t_0^{rest}$ are mainly less than $\pm 0.3$, $\pm 0.4$, $\pm 0.2$,
$\pm 0.2$, $\pm 0.4$ and $\pm 4$, respectively.

  \subsection{The Constraints on Cosmology}

\begin{figure}[htbp]
  \includegraphics[scale = 0.36]{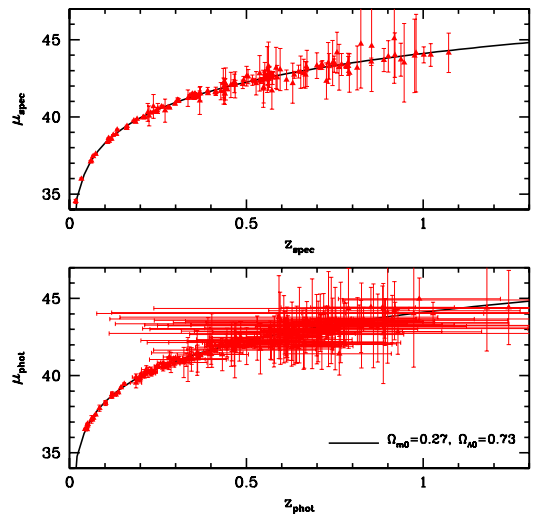}
  \caption{\label{fig:mu_z} The Hubble diagram for the
  $z_{spec}$ and $z_{phot}$ simulations. The data points
  on each figure are just one sixth of the whole data sets.}
  \end{figure}

  To establish the overall effect of the uncertainty from photometric
redshift for cosmological studies, we also generate 1000 SNeIa with
spectroscopic redshifts $z_{spec}$ with the LSST photometric error
$\sigma_{phot}$, i.e. we just fix the redshift and only model fit
the other five parameters. The Hubble diagram for the spectroscopic
and photometric cases are shown in Fig.\ref{fig:mu_z}. Only one
sixth of the whole data are shown on each figure.

  We use the MCMC approach to fit the cosmological parameters from the
two Hubble diagrams. Two cosmological scenarios are  considered,
that first one is $\rm \Lambda CDM$ with non-flat geometry and the
second is $\rm wCDM$ with the time-evolved equation of state for the
dark energy with $w(z)=w_0+w_1z/(1+z)$.

\begin{figure}[htbp]
  \includegraphics[scale = 0.36]{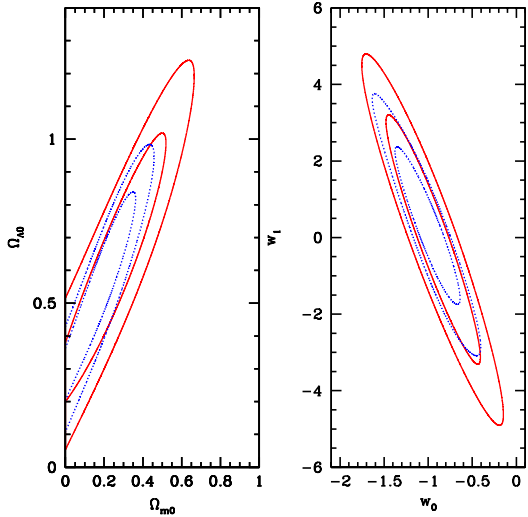}
  \caption{\label{fig:ml_w0w1} The contour maps for
  $\Omega_{m0}$ vs. $\Omega_{\Lambda 0}$ (left) and $w_0$ vs. $w_1$ (right).
  The $1\sigma$ and $2\sigma$ errors    are shown. The red
  solid and blue dotted contours are for photometric and
  spectroscopic redshift simulations respectively.}
  \end{figure}

 In Fig.\ref{fig:ml_w0w1}, we show the contour maps of $\Omega_{m0}$
vs. $\Omega_{\Lambda 0}$ and $w_0$ vs. $w_1$ with$1\sigma$ and
$2\sigma$ errors. As can be seen, the $1\sigma$ contours using
$z_{phot}$ and its error in cosmological parameter fits nearly
overlap with the $2\sigma$ contours of the case where redshift is
known precisely using$z_{spec}$. Also, we find little deviation for
the directions of the main axis of the contours for the two cases.
Thus, for a survey such as those planned for LSST, we effectively
find a factor of $\sim$ 2 degradation in parameter uncertainties
when using the SNIa sample with only photometric redshifts compared
with one with also spectroscopic redshifts.

In terms of the dark energy figure of merit that involves the
inverse area of the $w_0$ vs. $w_1$ ellipse, photometric SNe samples
lead to a factor of 4 degradation compared to spectroscopic sample.
This difference, however, is likely to be a minor issue: compared
with the planned SNeIa surveys which will involve spectroscopic
measurements of a few thousand SNe per year, photometric only
surveys such as the one with LSST will produce a sample of a few
hundred thousand SNe. Moreover, we have to note that for a real
survey the sample is always magnitude-limited, so that some $z>1.1$
objects could contaminate the $z<1.1$ sample and lead to a bias in
the fitting results. Given that we cannot quantify this bias
fraction, we don't include this effect in our analysis. In an
upcoming paper, we hope to implement a new technique to account for
such biases in large SNe samples.

  \section{Removing The Contamination from SNIb/c and SNII}

In a pure photometric survey such as the one with LSST without
spectroscopic measurements to identify if each of the SN is Type Ia
or not, in addition to the error in the measurement of redshift, the
photometric SNe samples would also be contaminated by core-collapse
supernovae. Here we consider the contamination from Type Ib/c
(SNIb/c) and Type II supernova (SNII).

  \subsection{Estimating The Contamination}

  To estimate the level of contamination, we create mock samples of
light curve data for SNIb/c and SNII. The spectral templates for
SNIb/c is from \citet{levan05}, and for SNII we use the templates of
SNIIP and SNIIL given by \cite{gilliland99} and
\cite{baron04}.\footnote{http://supernova.lbl.gov/$\sim$nugent/nugent\_templates.html}
Here, for simplicity, we consider primarily the
SNIIP and SNIIL. The SNIIn which have ``unusual'' progenitors \citep{mobberly07}
may be an important contamination for the SNIa \citep{poznanski071}.
We may discuss these objects in future work. We set the
percentage of SNIIP and SNIIL are $50\%$ and $50\%$ respectively for
the SNII sample. The mock flux data of the SNIb/c and SNII samples
are generated with the same procedure as those used to generate the
SNIa mock data in \S~2.

Since the core-collapse supernovae are intrinsically fainter
than the SNeIa and have no magnitude-phase relation, we take the
absolute peak magnitude from \cite{richardson02} and the Gaussian
distribution with $\bar{z}=0.4$ for SNIb/c and SNII, and then set
$s=1$ when mimicking their B-band peak magnitude. Also, given that
the core-collapse supernovae are usually found in star forming
regions, they are expected to suffer more extinction from the host
galaxy. We set $-0.2<E_{B-V}^{host}<0.4$ with
$\overline{E_{B-V}^{host}}=0.0$ and $\sigma_{E}=0.3$. Once
simulated,  we continue to use the SNIa SED light curves to fit them
just as we did in \S~3.

  \begin{figure}[htbp]
  \includegraphics[scale = 0.36]{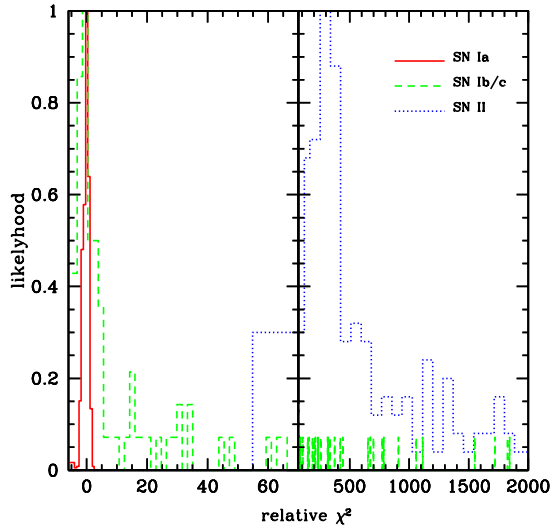}
  \caption{\label{fig:chi2} The distribution of relative $\chi^2$
  for the SNIa, SNIb/c and SNII samples when analyzed in all cases with Ia SED based light curves.
  The $\chi^2$ distribution of SNII data are significantly
  than the same for SNIa. This allows Type II SNe that are contaminating Ia samples to be easily
  distinguished.  However, the peak of the $\chi^2$ distribution for the SNIb/c overlaps with the same for SNIa and we
  find  SNeIb/c to be the main contaminants for photometrically selected Ia samples for cosmological measurements.}
  \end{figure}

The distribution of the difference of $\chi^2_{min}$ (i.e. relative
$\chi^2$) for the SNIa, SNIb/c and SNII are shown in
Fig.\ref{fig:chi2}. We find that when fitted with Ia SED light
curves, the $\chi^2$ values for SNeII are so large large values that
they are easily distinguished from the SNIa even with photometric
data alone. However, for SNeIb/c the $\chi^2$ peak overlaps with
that of the SNeIa, so they are the primary contamination to the
total sample.

To obtain a less-contaminated sample, as a first cut we note that
the $\chi^2$ distribution of the SNeIb/c has a long tail which can
extend to tens of thousands, and some of the SNeIb/c can be removed
by an overall restriction on the $\chi^2$ values in the fitting to
SNIa light curve template. If the selection is restricted to
$\chi_{rel}^2<20$, keeping all real Type Ia's, this results in a
removal of about $40\%$ of the  SNeIb/c's. Since the ratio of the
rate of SNIa to SNIb/c out to $z \sim 1$ is about 10 to 7
\citep{calura06,dePlaa07,sato07,poznanski072,eldridge08,smartt09,georgy09},
we expect about $250$ SNeIb/c to remain and contaminate a sample
that contains $1000$ SNeIa selected photometrically. Note that here
we just use a simplified assumption with the ratio of the rate of
SNIa to SNIb/c to be redshift independent. In the next subsection we
discuss a statistical method to further reduce the contamination of
Ib/c's during model fits to the total sample.

  \subsection{The Bayesian Statistical Method}

  We employ the Bayesian estimation method proposed by \citet{press96}
and \citet{kunz07} to further reduce the contamination from SNIb/c.
We note that our proposed statistical method cannot distinguish each
SNIb/c from a Type Ia individually, but statistically it reduces the
overall contamination and the associated bias in cosmological
parameters. As we illustrate here, the same method also allows us to
jointly estimate the fraction of the SNIa (or Ib/c's) within the
whole supernovae sample used for cosmology.

     We take the case that the observational sample  of supposedly
Type Ia's ${\bf D}$ contains a mixture of true SNeIa data ${\bf
D_1}$ and SNeIb/c's ${\bf D_0}$ which mimicks Ia. We define a vector
${\bf v}$ of length the total number of SNe $N$ with the value of
$v_i$ taking either $1$ or $0$ if $D_i$ is or is not a SNIa. We also
define a quantity $p$ to account for the total fraction of the true
SNeIa in the total SNe data set ${\bf D}$. Using ${\bf v}$ and $p$,
the posterior probability can be written as
  \ba
  P(\theta|{\bf D}) &=& \sum_{{p,\bf v}}P(\theta, {\bf v}, p|{\bf D}) \label{eq:bys_dand}\\
                    &\propto& \sum_{p,{\bf v}}\mathcal{L}({\bf D}|\theta,{\bf
                    v},p)P(\theta, {\bf v}, p)  \label{eq:bys_t1}\\
                    &\propto& \sum_{p,{\bf v}}\mathcal{L}({\bf D}|\theta,{\bf
                    v},p)P(p)P(\theta|p)P({\bf v}|\theta, p), \label{eq:bys_t2}
  \ea
  In Eq.~(\ref{eq:bys_dand}), the sum over $p$ will be the integration if the value of $p$
  is continuous, and the sum of ${\bf v}$ goes through all $2^N$ possible values of ${\bf v}$.
  Eq.~(\ref{eq:bys_t1}) is derived from the Bayes theorem, and
  $\mathcal{L}({\bf D}|\theta, {\bf v},p)$ is the likelihood. The $P(\theta|p)$ in
  Eq.~(\ref{eq:bys_t2}) can be reduce to $P(\theta)$ since there is no reason to believe
  the parameter $p$ affect the cosmological evolution of the Universe (it is not a cosmological parameter).

  Thus, we can simplify to
  \bq
  P(\theta|{\bf D}) \propto P(\theta)\sum_p P(p)\sum_{\bf v}
                    \mathcal{L}({\bf D}|\theta,{\bf v},p)P({\bf v}|\theta,p).
  \eq
  For any value of $p$, $P({\bf v}|\theta,p)$ is 0
  when $v_i$ involving the $i$th datum is a SNIb/c. When normalized,
  $P(v_i=1|{\theta},p)=p$ and $P(v_i=0|\theta,p)=1-p$. Therefore, we find
  \bq
  P(\theta|{\bf D})\propto P(\theta)\sum_p P(p)\sum_{\bf v}
                   \Big[\prod_{v_i=1}\mathcal{L}_i^1p\prod_{v_i=0}\mathcal{L}_i^0(1-p)\Big].
  \label{eq:bys_t3}
  \eq
  Here $\mathcal{L}_i^1$ is the likelihood that the $i$th SN is a Ia and this is taken to be
  \bq
  \mathcal{L}_i^1 = \frac{1}{\sqrt{2\pi}\sigma_i}e^{-\chi_i^2/2},
  \eq
  where $\sigma_i$ is the error and $\chi_i^2=(\mu_i^{obs}-\mu_i^{th})^2/\sigma_i^2$,
  and the $\mu_i^{obs}$ and $\mu_i^{th}$ are the observational and theoretical distance
  modulus respectively.

  The $\mathcal{L}_i^{0}$ is the likelihood for the SNIb/c samples though
  we don't know a priori the exact distribution. We use two parameters $b$ and $\sigma_0$ belonging to
  the parameter set $\theta$ to describe $P_i^0$ as
  \bq
  \mathcal{L}_i^0 = \frac{1}{\sqrt{2\pi}\sigma_0}e^{-\chi_0^2/2},
  \eq
  where $\chi_0^2=(\mu_i^{obs}-\mu_i^{th}-b)^2/\sigma_0^2$.

  We can simplify Eq.(\ref{eq:bys_t3}) further by noting that the $2^N$
  summation term can be written as the product of $N$ terms. We finally get
  \bq
  P(\theta|{\bf D})\propto P(\theta)\sum_p P(p)\prod_N
                   \Big[\mathcal{L}_i^1 p + \mathcal{L}_i^0(1-p)\Big].
  \eq
  The sum over $p$ is easily performed with MCMC runs, and we assume $P(p)$ is a
  uniform distribution. When analyzing our mock samples we take the ranges for $p$, $b$ and
  $\sigma_0$ of $p\in(0.5, 1)$, $b\in(-20, 20)$, $1/\sigma_0\in(0,
  1000)$.

  \subsection{The Results}

  Extending the discussion in \S 4.1, we add 250 SNIb/c data with
$\chi_{rel}^2<20$ to the 1000 SNIa data, and extract cosmological
constraints on the time-evolving equation of state of dark energy
with an analysis which implements the Bayesian estimation method
described above, in addition to a method where all data are analyzed
with a MCMC run without making an attempt to account for Ib/c
contamination to the total sample.

  \begin{figure}[htbp]
  \includegraphics[scale = 0.36]{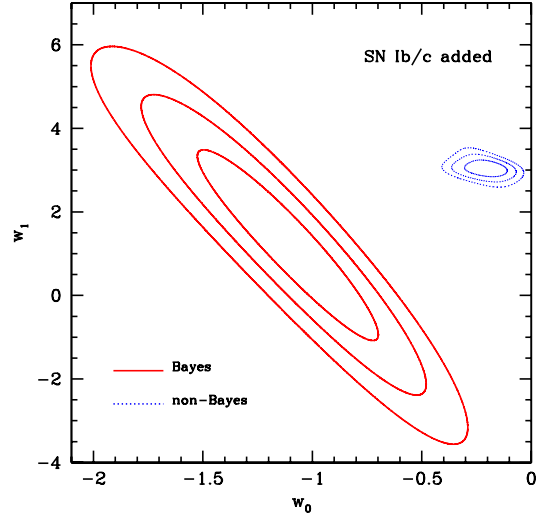}
  \caption{\label{fig:bys} The contour maps for $w_0$ vs. $w_1$
  with and without Bayesain estimation. The $1\sigma$,
  $2\sigma$ and $3\sigma$ errors are shown.}
  \end{figure}

  \begin{figure}[htbp]
  \includegraphics[scale = 0.36]{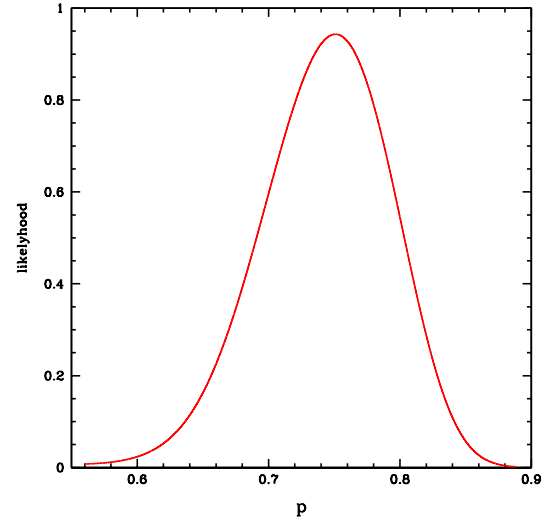}
  \caption{\label{fig:p} The PDF of the fraction p for the SNeIa in the mixed sample of Ia's and Ib/c's.
  The fiducial value is $0.8$ while the fitting result based on the technique outlined in \S~4.2
is   $p=0.75_{-0.06}^{+0.05}$.}
  \end{figure}

  We show the contour maps of $w_0$ vs. $w_1$ in Fig.\ref{fig:bys}.
The red solid and blue dotted contours are the fitting results with
and without Bayesian estimation, respectively. As can be seen in
Fig.\ref{fig:bys}, the constraint on $w_0$ and $w_1$ is completely
wrong when we ignore the Ib/c contamination in the total sample and
just do direct fitting to the Hubble diagram. This is caused by the
large differences of the distribution between the SNIa and the
SNIb/c data and  MCMC chains are easily trapped in a wrong
likelihood value. The $\chi^2$ value for the overall best-fit in
this case is also very large reaching as high as $2000$.

  When we implement the Bayesian estimation method, the result is
improved significantly. Although there is a difference between the
best fit and the actual (fiducial) value, the fiducial value of the
cosmological parameter set $(w_0,w_1)=(-1,0)$ lies safely within the
$2\sigma$ contour around the best fit. Also, as discussed, we also
jointly estimate the fraction of SNeIa in the total data set. We
plot the likelihood for $P(p)$ in Fig.\ref{fig:p}. The fraction of
the SNIa in this particular mock data should be  $80\%$ while the
fitting leads to the result of $p=0.75_{-0.06}^{+0.05}$ with errors
at 1$\sigma$. While there still remains a bias associated with the
contaminating Type Ib/c's, we have reduced this bias to the level of
a few percent. Also, if the distributions of $b$ and $\sigma_0$
are better measured in the future, this method would get better
typing result.

We believe the Bayesian estimation method provides a useful
statistical tool to reduce contamination and to evaluate the
fraction of the contaminating supernovae in the LSST photometric
SNIa survey. Of course, to identify whether each individual SN is a
core-collapse one or a Ia, the method discussed above is inadequate,
but the Bayesian statistical analysis can be a valuable guide for
further advanced study
\citep{poznanski02,gal-yam04,johnson06,poznanski071,kuznetsova07}.

  \section{Summary}

  In this paper, we explore the ability to determine the redshift and
the other parameters useful to construct the Hubble diagram with
light curve for the LSST SNIa photometric measurements. Using a SNIa
SED template and the expected photometric error of the LSST, we
first simulate the observed flux data of 1000 SNIa in each of 5 LSST
filters, and then apply a MCMC technique to fit the redshift,
stretch factor, apparent magnitude and the phase of the SNIa, among
others. We find that when $z<0.2$, these parameters can be
determined accurately at a level comparable to the case where
spectroscopic redshift is known. At higher redshifts, the
uncertainty in photometric redshift goes up quickly since
$\sigma_{phot}\sim 10^{0.4m}$, but the photometric data is still
very useful when  $0.2 < z < 0.5$. To illustrate the effect of the
uncertainty of the photometric redshift on the fitting of the
cosmological parameters, we also extract cosmological constraints
using parameters of the SNIa light curves with and without
spectroscopic redshifts. Using the fitting results of the two cases,
we constrain the cosmological parameters for $\Omega_{m0}$ and
$\Omega_{\Lambda 0}$ in the $\rm \Lambda CDM$ model and $w_0$ and
$w_1$ in the time-evolved $\rm wCDM$ model. We find that for the
same number of SNIa data, the cosmology fitting with only the
photometric data leads to a factor of 2 degradation in error of
cosmological parameters or a factor of 4 in the figure of merit of
dark energy equation of state (i.e. the inverse area of the
$w_0-w_1$ ellipse) compared with the case of fitting with
spectroscopic data. However, as the number of photometric-only data
far exceeds that with spectroscopic data, the overall statistical
uncertainty in the former would still be smaller.

  Finally, we discuss the contamination on the SNIa data from
core-collapse supernovae involving types II and Ib/c,  and the
feasibility of using a Bayesian estimation statistical method to
reduce the overall contamination. Similar to SNIa mock samples, we
generate the mock flux data for the SNIb/c and SNII based on their
spectral templates, and use the SNIa fitting process to fit them. We
find that the SNeII are easily distinguished from SNIa because there
is an apparent mismatch (large $\chi^2$) when fitting with the SNIa
templates. However, this is not the case for Type Ib/c's. The peak
of its $\chi^2$ distribution is overlapping with that of the SNIa
and present a significant contamination of any photometric selected
supposedly SNIa samples, even if a conservative cut is applied in
the $\chi^2$ values for selection. To further account for this
contamination, at least statistically when doing cosmological model
fits, we employ Bayes theorem. Our suggested method could reduce the
contamination down to a few percent level, leading to estimates of
cosmological parameters that are biased within 1$\sigma$ errors. The
method also establishes the  fraction true SNIa in the total
photometric SNe data set. Nevertheless, we must note that this
method cannot distinguish if an individual SN is whether Type Ia or
not. We will need an extended analysis complemented with additional
observations if we are required to recognize the type of individual
SNe.

  \begin{acknowledgments}
This research is supported by the NSF under CAREER AST-0645427 at UCI, by the NSFC under the Distinguished Young
Scholar Grant 10525314, the Key Project Grant 10533010, by the
Chinese Academy of Sciences under grant KJCX3-SYW-N2, and by the
Ministry of Science and Technology of China under the National Basic Science
program (project 973) grant 2007CB815401.
X.C. also acknowledges the hospitality of the Moore Center of Theoretical
Cosmology and Physics at Caltech, where part of this research is performed.
  \end{acknowledgments}

  %\appendix

  %\section{}

  \end{document}